\documentclass[12pt,a4paper]{article}
\usepackage[
    backend=bibtex, 
    style=numeric-comp, 
    sorting=none, 
    sortlocale=auto,
    maxbibnames=3, 
    minnames=3, 
    giveninits=true, 
    url=false, 
    doi=true, 
    eprint=false,
    isbn=false, 
    hyperref=true, 
    backref=false, 
    block=none
]{biblatex}

\usepackage{setspace}
\usepackage[top=1in, bottom=1.25in, width=180mm]{geometry}
\usepackage{graphicx}
\usepackage{dcolumn}
\usepackage{bm}
\usepackage[normalem]{ulem}
\usepackage[utf8]{inputenc}
\usepackage{newtxtext}
\usepackage{newtxmath}
\usepackage[T1]{fontenc}
\usepackage{textcomp}
\usepackage{nicefrac}
\usepackage{xcolor}
\usepackage{upgreek}
\usepackage{hyperref}
\hypersetup{colorlinks=true, citecolor=blue, linkcolor=blue, urlcolor=blue}
\usepackage{ragged2e}
\usepackage{physics}

\usepackage[activate={true,nocompatibility},final,tracking=true,kerning=true,spacing=true,factor=1100,stretch=10,shrink=10]{microtype}
\SetProtrusion{encoding={*},family={*},series={*},size={6,7}}
{1={ ,750},2={ ,500},3={ ,500},4={ ,500},5={ ,500},
6={ ,500},7={ ,600},8={ ,500},9={ ,500},0={ ,500}}
\SetExtraKerning[unit=space]
{encoding={*}, family={*}, series={*}, size={footnotesize,small,normalsize}}
{\textendash={400,400}, 
    "28={150,100},
    "29={100,150},
    \textquotedblleft={ ,150},
\textquotedblright={150, }}
\SetTracking{encoding={*}, shape=sc}{10}

\DeclareMathAlphabet{\pazocal}{OMS}{zplm}{m}{n}

\newcommand{\momipy}{\texttt{momipy}}
\newcommand{\mos}{$\text{2H-MoS}_2$}
\newcommand{\mosbl}{$\text{BL-MoS}_2$}
\newcommand{\figlabel}{Figure~}

\usepackage{titlesec}
\titleformat{\section}{\sc\large\sffamily\bfseries}{}{0.5em}{}[{}]
\titleformat{\subsection}{\sc\sffamily\bfseries}{}{0.5em}{}[{}]

\usepackage{caption}
\DeclareCaptionJustification{raggedright}{\justifying}

\usepackage{authblk}

\DeclareFieldFormat{journaltitle}{\textit{#1}}
\DeclareFieldFormat[inbook,thesis]{title}{`#1`} 
\DeclareFieldFormat[article, misc, thesis, book]{title}{`#1`} 
\DeclareFieldFormat[article]{volume}{\textbf{#1}\addcolon\space}
\DeclareFieldFormat{pages}{#1}

\begin{document}

\title{\textbf{\momipy: A time-resolved momentum microscopy data analysis workflow}}

\author[1]{Paul Werner}
\author[1]{Wiebke Bennecke}
\author[1]{Jan Philipp Bange}
\author[1]{Junde Liu}
\author[1]{Daniel Steil}
\author[2,3]{Marcel Reutzel}
\author[1,4]{G.~S.~Matthijs~Jansen\thanks{jansengsm@leidenuniv.nl}}
\author[1,5]{Stefan Mathias\thanks{smathias@uni-goettingen.de}}

\affil[1]{I.~Physikalisches Institut, Georg-August-Universit\"at G\"ottingen, Friedrich-Hund-Platz 1, 37077 G\"ottingen, Germany}
\affil[2]{Fachbereich Physik, Philipps-Universität Marburg, 35032 Marburg, Germany}
\affil[3]{mar.quest $\vert$ Marburg Center for Quantum Materials and Sustainable Technologies, 35032 Marburg, Germany}
\affil[4]{Leiden Institute of Physics, Leiden University, Niels Bohrweg 2, 2333 CA Leiden, The Netherlands}
\affil[5]{International Center for Advanced Studies of Energy Conversion (ICASEC), University of G\"ottingen, G\"ottingen, Germany}

\date{\today}

\maketitle

\begin{abstract}
    Photoemission momentum microscopy is an emerging approach to photoelectron spectroscopy that is transforming electronic band structure mapping at both laboratory-based and large-scale facilities.
    Its extension to time-resolved measurements has recently enabled major advances, ranging from the femtosecond imaging of exciton wavefunctions to dark-field photoelectron microscopy of light-induced dynamics and strong light-matter interactions.
    However, the rapid acquisition of high-dimensional data sets (including electron in-plane momenta $k_x$ and $k_y$, kinetic energy $E_{\rm kin}$, time $\Delta t$, and more) poses substantial challenges for data handling and analysis.
    A typical workflow encompasses data acquisition, exploratory visualization, calibration and targeted feature extraction, with data management ideally guided by FAIR principles.
    Here, we present \momipy, a Python package that streamlines the momentum microscopy data analysis workflow efficiently and at low computational cost.
    Among other features, \momipy{} provides interactive 4D data visualization, energy and momentum calibration, space-charge correction, and a broad selection of multidimensional data analysis routines, including binning, momentum-map extraction and time-, energy- and momentum-resolved model fitting.
    \momipy{} is accompanied by a selection of interactive example scripts, supporting adoption of \momipy{} by both experienced researchers and users new to momentum microscopy.
\end{abstract}
\newpage

%%%%%%%%%%%%%%%%%%% Section 1: Introduction
\section{Introduction}
Time- and angle-resolved photoelectron spectroscopy (tr-ARPES) provides unrivaled access to the electronic structure of materials, combining full momentum information with the ability to map ultrafast dynamics with femtosecond time resolution~\cite{boschini_time-resolved_2024}.
Recently, the extension of this technique by so-called momentum microscopes in combination with short-pulse extreme ultraviolet light sources has enabled a series of landmark discoveries, ranging from the tracing of exciton dynamics in 2D semiconductors~\cite{Madeo20sci, Schmitt22nat, kunin23prl} and organic semiconductors~\cite{Neef2023, Wallauer20sci, Bennecke24natcom, Bennecke25Nat.Phys., baumgartner_multiplex_2023} to the observation of Floquet-Bloch states~\cite{Merboldt25NatPhys} and Landau–Zener–Majorana transitions in graphene~\cite{eggers_subcycle_2026}.
The driving force behind these discoveries was the development of dedicated ultrafast time of flight momentum microscopy setups at several laboratories~\cite{Keunecke20timeresolved, bennecke_table-top_2026, Maklar20revsciint, fragkos_time-_2025, Wallauer20sci, heber_multispectral_2022, Allison2025_aplphot, Karni2023_advmat, Schiller2025_scirep, haag_time-resolved_2019}, including ours, that combine momentum microscopy with high-repetition-rate high-harmonic-generation light sources~\cite{corder_ultrafast_2018, saule_high-flux_2019, Eich14jesrp, chiang_boosting_2015}. Time-resolved momentum microscopy is also becoming increasingly available at facility beamlines, including free-electron lasers~\cite{eli_alps, kutnyakhov_time-_2020, Dendzik2020, Grychtol_SXP_2022, Tkach2026_multimode}.
By joining an electron microscope lens column with momentum-resolved detection and time of flight energy resolution, momentum microscopy can measure multidimensional datasets spanning in-plane momentum coordinates, kinetic energy, pump-probe delay and additional parameters such as temperature, photon energy, polarization, spin, or geometry with unprecedented efficiency~\cite{medjanik_direct_2017, kroemker08rev_sci_int, schonhense_space-_2015, tusche_spin_2015}.
While conventional ARPES can in principle access similar multidimensional information, momentum microscopy achieves this with vastly superior data collection rates due to its non-scanning measurement principle, enabling measurements that would be prohibitively time-consuming or impractical with traditional ARPES setups.

\begin{figure}[!pb]
    \centering
    \includegraphics{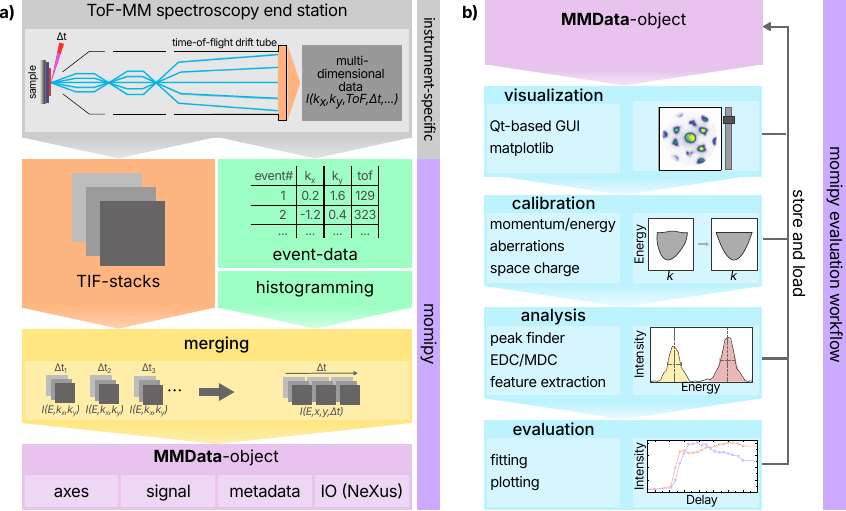}
    \caption{
        Overview of the workflow using \momipy{} for time of flight (ToF) momentum microscopy (ToF-MM), from data generation and preprocessing (a) to data analysis (b). 
        A typical ToF-MM end station produces multidimensional datasets. 
        These raw data is ingested into the workflow in either TIF stack (orange) or single-electron event-based format (green) and histogrammed. 
        Merging (yellow) then combines the multidimensional datasets into a single working dataset, which is then processed in the MMData format. 
        The MMData object stores the axes and signal, as well as metadata, and provides important input/output operations. 
        From that, the data can be calibrated and corrected, analyzed and/or visualized using the provided tools or other resources.
    }
    \label{fig:fig_overview}
\end{figure}

Momentum microscopy at its core measures the photoemission intensity $I$ depending on both in-plane momenta $k_x, k_y$ and kinetic energy $E_\text{kin}$ of the photoemitted electrons.
Thus, a single static measurement produces an (at least) three-dimensional dataset containing $I(k_x, k_y, E_\text{kin})$ data.
This might be recorded and stored directly in histogram form. Alternatively, a data stream might be recorded where a list of detected electrons is stored along with their in-plane momenta, kinetic energy and detection timestamps as an event list (see \figlabel\ref{fig:fig_overview}a).
While the latter approach enables further data preprocessing opportunities, the starting point for data analysis is in almost all cases a three- (or higher) dimensional histogram.
With pump-probe delay $\Delta t$ introducing an additional temporal axis, time-resolved momentum microscopy inherently possesses a high data complexity.
Further experimental degrees of freedom such as temperature, pump/probe polarization, photon energy or spin dependence add yet more axes to the multidimensional dataset.
If not handled carefully, this can rapidly increase the data file sizes and cause the evaluation procedure to exceed the computational resources available.
Hence, a crucial challenge is the distillation of key information from the collected raw data in a systematic, time- and resource-efficient manner.

Beyond the handling of large multidimensional datasets, a second challenge lies in disentangling relevant features of the photoemission spectrum from instrumental measurement artifacts.
Despite its power, momentum microscopy is prone to instrument-induced distortions, mainly from space charge and the lens system, that are significantly more pronounced than in conventional time-resolved ARPES experiments~\cite{Maklar20revsciint}.
Aberrations of the electrostatic lenses can lead to anisotropic and energy-dependent momentum scaling~\cite{tromp11_ultramic, Tusche2019_ultramic, Tkach2025_ultramic, xian2019_ultramic}, which, if neglected, results in incorrectly assigned momentum axes that propagate into all subsequent analysis steps.
Additionally, the Coulomb interaction between the photoemitted electrons in the lens column of the microscope can cause momentum-dependent energy shifts and broadening of spectral features~\cite{schoenhense_space_charge, passlack_space_2006, Oloff16jap, Hellmann09prb}.
These so-called space-charge effects can lead to an incorrect energy assignment of spectral features, and, more critically, could be misinterpreted as a material response rather than an artifact.
These two examples highlight the importance of detecting such distortions and properly accounting for them before interpreting the data.

Despite the ongoing establishment of momentum microscopy in labs worldwide, adequate software infrastructure for data handling and analysis is still lacking.
While several tools exist that have been developed for conventional 2D ARPES workflows~\cite{stansburypyarpes, PIVA, erlab-py, peaks}, these are poorly suited to the demands of time-resolved momentum microscopy data.
They lack automation features for reproducible data evaluation or momentum-microscopy-specific tooling.
This applies specifically to the identification and handling of instrument-induced distortion.
Several correction methods have been proposed~\cite{Schoenhense2015_ultramic, xian2019_ultramic, Tusche2019_ultramic, MPES, schoenhense_space_charge}, but significant improvements are still needed regarding development, sophistication, ease of use, and the quantification of how these corrections impact the data.
Consequently, practitioners typically develop their own custom analysis code: a time-consuming process that could hinder reproducibility and complicate the adoption of FAIR-compliant data management practices, i.e. Findable, Accessible, Interoperable and Reusable data sets, for example built using the NeXus data format~\cite{nexusformat}.
A notable contribution is the {\tt MPES} package~\cite{MPES}, which combines the processing of event-based momentum microscopy data with analysis and visualization capabilities.
However, the workflow of {\tt MPES} is specifically tailored to event-based collection schemes, focusing on the transition from event data to evaluation-ready histogram datasets, which are subsequently analyzed by custom evaluation procedures.

We developed \momipy{}, a Python library with specialized tooling for multidimensional photoemission spectroscopy data in order to provide scientists with a batteries-included, fast-paced evaluation platform that complies with the FAIR principles. 
Complementing {\tt MPES}~\cite{MPES}, \momipy{} focuses on the handling of multidimensional histogram datasets and strives to streamline the workflow from raw histogram-based data to final publication-grade figures (see \figlabel\ref{fig:fig_overview}a). 
At its center lies a data class which holds multidimensional data, supports standard input-output (IO) operations, and is fully compatible with the HDF5 standard~\cite{HDF5} and NeXus format~\cite{nexusformat}. 
Additionally, \momipy{} supports the \texttt{NXmpes} specification, which is a standardized format for multidimensional photoemission experiments~\cite{nexusformat}. 
Surrounding the data class are multiple modules that are responsible for converting raw experiment data to the aforementioned data class format, data correction and general evaluation, as highlighted in \figlabel\ref{fig:fig_overview}b.
Furthermore, a cross-platform 3D/4D data viewer is included which allows interactive visualization of multidimensional data (see \figlabel\ref{fig:fig_viewer}).

We present the capabilities of \momipy{} by first describing its architecture and core principles. 
Following that, we highlight the features of \momipy{} using an exemplary end-to-end workflow, from acquired raw data to analysis results.

\section{Architecture and core functionality}

To be accessible by a broad audience, \momipy{} is built on top of Python's rich data science ecosystem. 
It uses \texttt{numpy}~\cite{numpy}, \texttt{scipy}~\cite{scipy}, \texttt{h5py}~\cite{h5py}, \texttt{matplotlib}~\cite{matplotlib}, \texttt{scikit-image}~\cite{scikit-image} and \texttt{lmfit}~\cite{lmfit} as core dependencies. 
Compatibility with the NeXus format is ensured via the \texttt{nexusformat} package~\cite{nexusformat}. 
The base of \momipy{} is its data class {\tt MMData}, which is responsible for holding $n$-dimensional data, together with axes and metadata. 
It is built using the NeXus-compatible data types from \texttt{nexusformat} and can easily be imported from or exported to NeXus files to promote shareability and access to the data by other tools. 
Since the NeXus data format is built on top of the well-established HDF5 standard, advanced features such as data compression, lazy-loading, and automatic chunking are inherently supported (see Methods for more details). 
\momipy{} supports multiple ways of creating an \texttt{MMData} object. 
Beyond directly loading from a NeXus file, it can be created by loading measured data, either histogram data in the form of TIF files or eventlist data via the provided merging and histogramming routines. 
Lastly, an \texttt{MMData} instance can be created manually. 
By design, the scope of the \texttt{MMData} class is focused on robust IO support, ensuring it remains highly accessible and straightforward to use.

All other functionality provided by \momipy{} is mostly independent of the \texttt{MMData} format. 
The features are split into individual modules. 
Key modules for data preprocessing are the {\tt corrector} and {\tt energycal} module, which facilitate the energy and momentum calibration and provide a comprehensive toolbox for identifying common distortions and applying corrections where physically justified.
Each correction step can be visualized to assess its suitability and its impact on the data.
While \momipy{} provides extensive preprocessing steps and corrections, it is the user's responsibility to ultimately assess whether their application is justified and sensible.
The most extensive of the modules is the {\tt analysis} module, which includes core methods for data evaluation.
To highlight a few functions, it provides a convenience function for extracting cuts through volumetric data, and the 2D \texttt{PeakFinder}, which provides reliable automated peak finding.
One example where this is particularly useful is the quick selection of regions of interest for creating energy distribution curves (EDCs). 
This example is discussed in further detail below.

\momipy{} also includes two tools featuring a graphical user interface (GUI): \textit{eventviewer} and \textit{energycal}.
The \textit{eventviewer} is a data visualization tool for 3D and 4D datasets, mainly geared towards time-resolved data.
Besides interactive data exploration, it supports selected analysis features, e.g. peak finding and EDC extraction.
The \textit{energycal} application is used for interactively determining energy calibrations, i.e. mappings from time of flight (ToF) to kinetic energy of the photoelectrons.
A key feature is support for automatic fitting of the energy calibration function.
All features used in \textit{energycal} are also available in the \texttt{energycal} module (see Methods and exemplary workflow section).

\subsection{Data visualization}
\begin{figure*}[!pt]
    \centering
    \includegraphics[width=\linewidth]{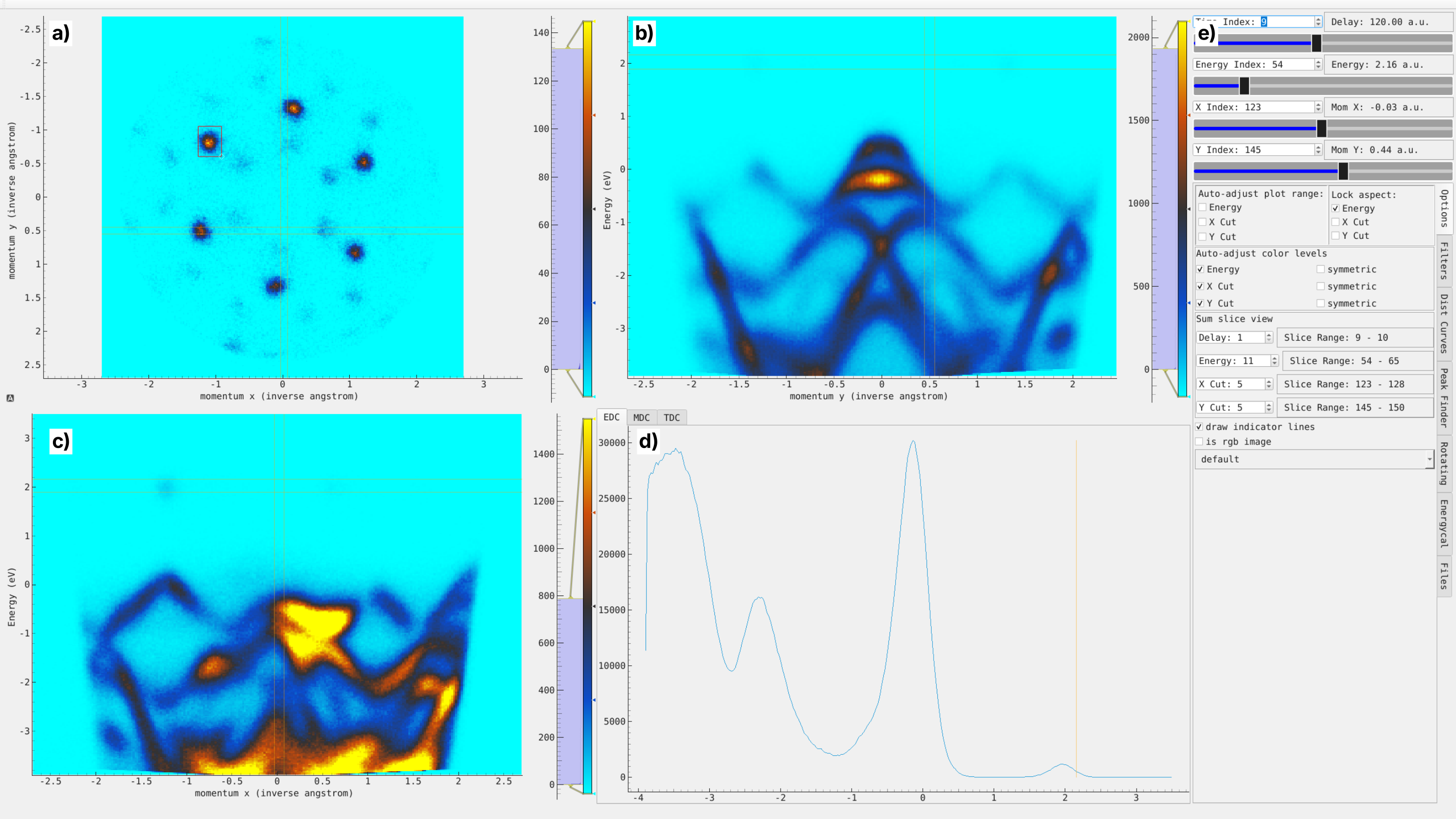}
    \caption{
        Example of the 4D viewer interface. 
        \textbf{a-c)} each correspond to a cut through the three-dimensional dataset. 
        The fourth dimension, in this case the delay time, can be adjusted by a slider in the control panel (e). 
        \textbf{d)} Panel for displaying energy, momentum or temporal distribution curves (EDC, MDC, TDC) through the data. 
        \textbf{e)} Control panel for adjusting the parameters of each slice.
    }
    \label{fig:fig_viewer}
\end{figure*}

While \momipy's analysis module already provides reasonably fast access to views into a dataset, sometimes real-time data exploration is required.
For this purpose, \momipy{} also provides a 3D/4D data viewer, called \textit{eventviewer}.
This is a GUI application which is built using Python bindings for the Qt6 framework.
Since the Qt6 framework is cross-platform, the GUI works on all major operating systems. 
The GUI allows for interactive exploration of datasets provided in either TIF, NeXus or eventlist format. 
These are then visualized in a viewer application, that features four main panels and a GUI control interface, as shown in \figlabel\ref{fig:fig_viewer}.
The main panels feature three orthogonal 2D cuts through the data (\figlabel\ref{fig:fig_viewer}a-c). For time-resolved momentum microscopy data these would be a constant-energy map and two energy-momentum maps. 
To control the view into the fourth dimension (e.g., pump-probe delay), a slider is used that selects a slice along this dimension and modifies the data displayed by the four views accordingly. 
The fourth panel (\figlabel\ref{fig:fig_viewer}d) features a line plot to visualize energy, momentum or temporal distribution curves, for which the integration region can be modified in the first view panel as well as the control panel.
Via the control panel (\figlabel\ref{fig:fig_viewer}e), multiple features can be accessed, such as filters for smoothing, dataset rotation, the distribution curve control and the \texttt{PeakFinder}.
Having access to an interactive viewer greatly reduces the entry barrier and enables users to quickly peek at datasets during measurements, which proved invaluable during measurement sessions.

\section{Exemplary end-to-end workflow}

In this section, we will discuss how \momipy{} can streamline the complete analysis workflow, i.e. from raw data to evaluation results, based on real-world raw data from our momentum microscope end station~\cite{Keunecke20timeresolved, werner2026_neq}. 
This example will cover a time-resolved dataset taken of a \mos~homobilayer, which is publicly available on \href{https://doi.org/10.25625/EKTPZZ  }{GRO.data}~\cite{EKTPZZ_2026}.
In this measurement, we recorded the dynamics of laser-induced excitons at the $K$ and $\Sigma$ valleys of the Brillouin zone. 
The data and worked examples in the form of Python scripts and Jupyter notebooks can be found in Ref.~\cite{momipy_gro_data}.

\subsection{Merging of raw data}
The starting point of our data analysis is the raw measurement data (see \figlabel\ref{fig:fig_overview}a). 
In the case of the homobilayer 2H-MoS$_2$, the data were recorded as a large collection of 3D TIF files: each TIF file holds the number of photoelectrons depending on ToF $\tau$ and the two in-plane momenta $k_x$ and $k_y$ as a dense array, integrated for a specified time span at a fixed pump-probe delay $\Delta t$. 
Since the presented measurement is a time-resolved measurement, we accumulate one such dataset for each pump-probe delay $\Delta t$. 
To reduce the influence of e.g. drift, long-term laser instability, or sample aging, we accumulate each time step for a rather short amount of time (typically around 3-10 minutes) and repeat the whole series multiple times to ensure good statistics. 
The first processing step of the evaluation workflow is therefore merging the individual TIF files that have been acquired for the same pump-probe delay.
This is handled using the \texttt{TifMerger} from the \texttt{merging} module. 
It provides methods for loading and merging the TIF files while applying energy and time delay corrections, as well as specialized plotting functions to highlight a few capabilities. 
After loading and merging, the data is written to a single \texttt{MMData} compatible NeXus file, which stores the merged, 4D time-resolved dataset.

\subsection{Data calibration and distortion correction}
\begin{figure*}[!pt]
    \centering
    \includegraphics[width=\textwidth]{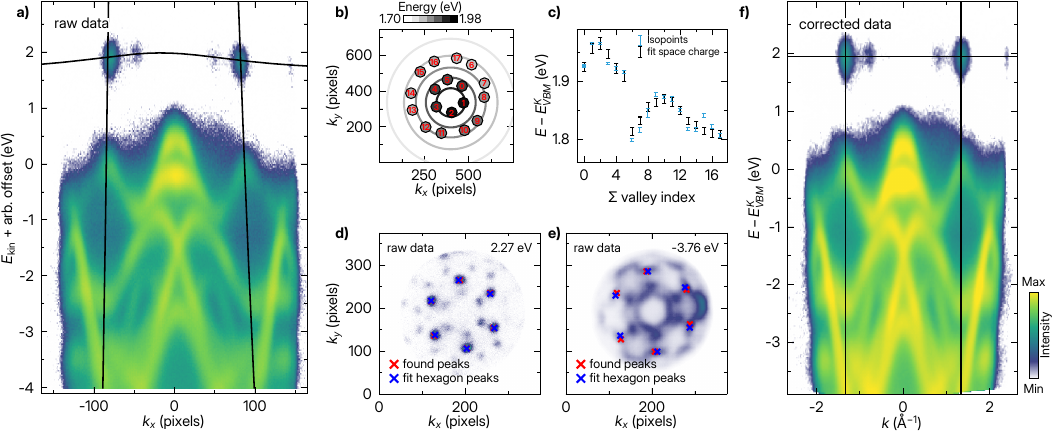}
    \caption{
        Data preprocessing including momentum calibration and space-charge correction.
        \textbf{a)} Energy-momentum cut through the raw data after preliminary energy calibration along the $K-\Gamma-K'$-direction. 
        Black lines are a guide to the eye for the determined space-charge induced energy shifts and lens aberration in momentum.
        \textbf{b)} Momentum map of the determined energies of the excited-state signal at the $\Sigma$ valleys in the first and second Brillouin zone.
        The gray contours indicate the result of a Lorentzian fit with constant background to the determined $\Sigma$ valley energies.
        \textbf{c)} Energy of each $\Sigma$ valley (blue) and the corresponding energy from the Lorentzian fit (black).
        \textbf{d--e)} Iso-energy momentum maps of the raw data centered at $E - E_\text{VBM}^K =2.27$~eV and $-3.76$~eV, respectively.
        Red crosses mark $K$ points determined by the {\tt PeakFinder}; blue crosses mark the ideal hexagon corners from the fit.
        \textbf{f)} Same energy-momentum cut as in \textbf{a)} after calibration and correction.
        Black lines serve as a guide to highlight the effects of the calibration.
    }
    \label{fig:fig_correction}
\end{figure*}

Before proceeding with the data analysis, the raw momentum microscopy data must be checked for distortions, and their energy and momentum axes need to be converted into physical units. 
\figlabel\ref{fig:fig_correction}a shows a cut through the raw photoemission data of a \mos~homobilayer, which is affected by both lens aberrations and space-charge induced distortions. 
The former cause an energy-dependent momentum scaling, the latter a momentum-dependent energy shift.

For the presented \mos{} measurements, the energy calibration and space-charge correction are combined into a single step. 
The energy calibration is first determined manually using the graphical interface of the \textit{energycal} tool and subsequently refined by a least-squares fit. 
The observed space-charge-induced energy shift is mostly delay-independent and is best seen for the momentum-dark excitons at the $\Sigma$ valleys of the conduction band, whose kinetic energies in the first and second (extended) Brillouin zone deviate substantially. 
To quantify this shift, \momipy{} extracts the energy position of these features and fits their momentum dependence with a 2D Lorentzian modeling the position-dependent space-charge-induced energy shift~\cite{schoenhense_space_charge}, as visualized in \figlabel\ref{fig:fig_correction}b,c. 
Subtracting this Lorentzian profile from the energy calibration corrects the space-charge shift and renders the calibration momentum-dependent (more details in Methods).

For the momentum correction and calibration, we determine the edges of the Brillouin zone at two different energies (centered around $E - E_\text{VBM}^K =2.27\,$eV and $E - E_\text{VBM}^K = -3.76\,$eV, see \figlabel\ref{fig:fig_correction}d,e). 
Using \momipy{}'s {\tt corrector} module, we map them onto a regular hexagon by an isotropic affine transformation, i.e. identical scaling in $x$ and $y$ without shearing. 
Since the center-to-corner distance of the hexagon corresponds to the known $\Gamma$K distance of \mos, this step corrects the energy-dependent distortions and calibrates the momentum axes simultaneously. 
The parameters are then linearly extrapolated from these two energy slices to all energies (see Methods for more details).

Finally, the calibrated and corrected data are resampled onto a uniform and equidistant grid with coordinates ($E$, $k_x$, $k_y$). 
To perform this data resampling, \momipy{} leverages robust interpolation routines from the \texttt{scipy} library~\cite{scipy}, supporting 0th-, 1st-, or 3rd-order spline interpolations, as well as Piecewise Cubic Hermite Interpolating Polynomial (PCHIP) and Akima formulations.

The difference between the merged, raw dataset and the corrected and calibrated data set is highlighted in \figlabel\ref{fig:fig_correction}a and f. 
The raw data shows a clear energy shift between the excited-state signatures of the $\Sigma$ exciton at $\approx2$ eV, which is removed by the space-charge correction. 
Moreover, the photoemission horizon of the uncorrected data is constant or even becomes smaller for increasing kinetic energy, whereas it displays the expected parabolic shape for the corrected data.

As a last step before proceeding with the evaluation, we normalize the time-resolved data such that intensity variations between the time steps that can occur due to differing integration times or changes in probe photon flux are corrected. 
This is achieved by summing the intensity for each dataset between $E - E_\mathrm{VBM}^K=-2.44\,\mathrm{eV} ~\mathrm{and}~ E - E_\mathrm{VBM}^K=-3.42\,$eV and normalizing such that this value is constant for all time steps. 
This normalized data is again saved as a NeXus file and is considered the working dataset for further evaluation.

\subsection{Data analysis}
\begin{figure*}[!pt]
    \centering
    \includegraphics{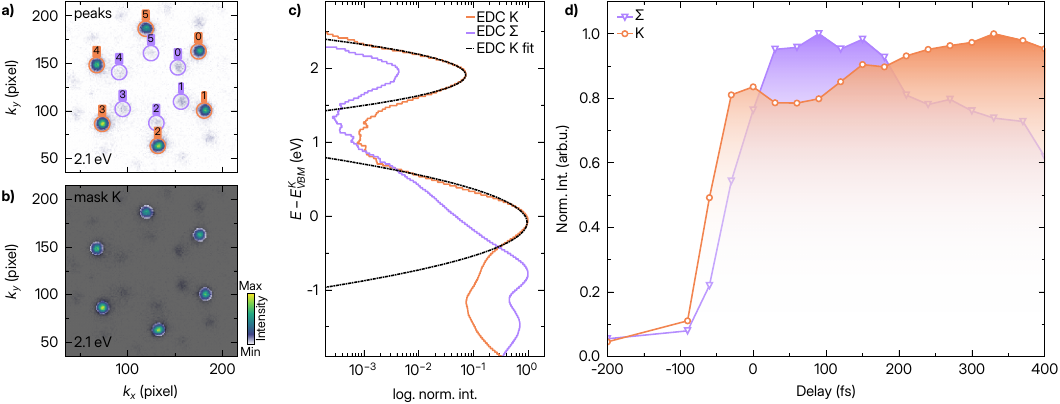}
    \caption{
        Overview of the \texttt{PeakFinder}, region-of-interest extraction and temporal analysis.
        (a) Momentum map of the excited state signal of \mosbl~with fitted peaks at the $K$ (orange) and $\Sigma$ (purple) valleys.
        (b) Momentum map from (a) with mask for spectrum extraction.
        (c) Resulting EDCs from the $K$ (orange) and $\Sigma$ (purple) valley masks. 
        Black line shows a fit to the EDCs at $K$.
        (d) Temporal evolution of the excited-state signal at the $K$ and $\Sigma$ valley extracted in the energy range from 2.4 to 1.7 eV.
    }
    \label{fig:fig_roi}
\end{figure*}

The next step in the analysis could be the extraction of excited-state features and the quantification of their time-dependent amplitude, mean energy, or momentum width. 
In this context, the \texttt{PeakFinder} is a powerful tool for automating data analysis. 
To highlight its capabilities, we will use it to extract EDCs at the $K$ valleys and fit them. 
First, we use the \texttt{PeakFinder} in the automatic mode to find the six brightest peaks in an iso-energy cut (momentum map), whose positions are then refined by fitting 2D Gaussians to the data. 
The momentum map and the peak positions are presented in \figlabel\ref{fig:fig_roi}a. 
Next, a mask is created from the extracted peak positions that excludes all the data except for a small region around the peaks. 
\figlabel\ref{fig:fig_roi}b shows the mask that is used to extract the photoemission yield within the region of interest. 
This mask is applied to all energy slices of the dataset and subsequently summed along the in-plane momentum directions. 
This gives us the desired EDC, shown in \figlabel\ref{fig:fig_roi}c, which can be evaluated further e.g, by fitting it with a suitable model. 
In time-resolved measurements, these EDCs can be extracted for all delay steps and filtered in energy to obtain time-resolved photoemission intensities, as shown in \figlabel\ref{fig:fig_roi}d. 
Here, we extracted the photoemission intensity of the $K$ and $\Sigma$ valleys for different pump-probe delays $\Delta t$.

\begin{figure*}[!pt]
    \centering
    \includegraphics[width=\linewidth]{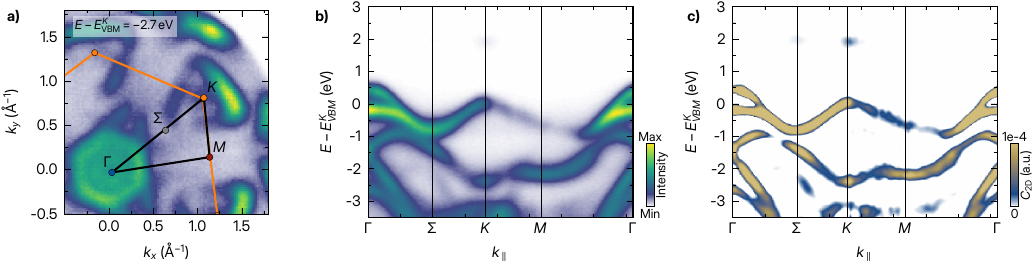}
    \caption{
        (a) Momentum map of the valence bands of \mosbl~at $E-E_\mathrm{VBM}^K=-2.7\,\mathrm{eV}$ and a pump-probe delay of $\Delta t=180\,\mathrm{fs}$. 
The orange line marks the first Brillouin zone, the $\Gamma$, $K$, $M$ and $\Sigma$ valleys are also highlighted.
        (b) Unfolded cut along $\Gamma$-$\Sigma$-$K$-$M$-$\Gamma$.
        (c) Same cut as (b) with 2D curvature filter (Ref.~\cite{zhang2011_curvature_filter}) applied to highlight bands.
    }
    \label{fig:fig_cut}
\end{figure*}

Another common procedure is taking cuts through the volumetric data. 
For this we can use the \texttt{volume\_cut} function from the analysis module. 
\figlabel\ref{fig:fig_cut}a shows a momentum map taken at $E - E_\mathrm{VBM}^K=-2.7\,\mathrm{eV}$ and a pump-probe delay of $\Delta t=180\,\mathrm{fs}$. 
The points mark the $\Gamma$ (black) point, the $\Sigma$ (gray) valley, the $K$ and $K$' (orange) valleys and the $M$ points (blue). 
We use the \texttt{volume\_cut} function to extract the cut through the volume from $\Gamma$ to $M$ via the $\Sigma$ and $K$ valleys.

The resulting unfolded energy-momentum cut is shown in \figlabel\ref{fig:fig_cut}b. 
We use a peak-corrected second derivative filter~\cite{zhang2011_curvature_filter} to enhance the visibility of the band structure, which is displayed in \figlabel\ref{fig:fig_cut}c.

\momipy{} has a built-in preset for plotting with matplotlib which applies sensible default options to the plots and supports different fonts. 
All plots shown in this manuscript are created using the built-in style. 
It also provides some useful matplotlib-related functions, like colormap creation and setting the lightness of a color.

\section{Discussion}
In summary, \momipy{} offers a solution to complicated and fragmented data processing pipelines in tr-ARPES by providing a unified framework. 
It provides access to merging routines, sophisticated momentum and energy calibration methods, distortion correction and interactive 4D data visualization, while also supporting compliance with FAIR principles. 
This enables efficient, reproducible and standardized analysis of multidimensional photoemission datasets, significantly lowering the barrier to entry and accelerating complex data evaluation for novices and experts alike. 
Since \momipy{} is built on the foundation of the rich scientific Python ecosystem, users are not locked into \momipy{}, but are free to extend its functionality where needed.

Looking forward, our efforts will focus on refining and validating the existing feature set. 
While core components such as the \texttt{MMData} class and binning routines are currently supported by unit tests, expanding test coverage across the entire package will ensure long-term stability and performance as the need for new features arises. 
Comprehensive testing also provides researchers with essential peace of mind, guaranteeing that critical scientific evaluations do not rely on untested or potentially erroneous code.

\section{Methods}
\setcounter{figure}{0}
\captionsetup[figure]{name={Method Fig.}, font={sf,small, stretch=1.2}, labelfont={bf}, , justification=raggedright,format=plain}

The photoemission data presented for the exemplary workflow originates from~\cite{werner2026_neq}. 
For further details, we refer to this publication.

A detailed description of the Göttingen in-house ultrafast momentum microscopy setup and its application to exfoliated van der Waals materials is provided in Refs.~\cite{Keunecke20timeresolved} and~\cite{Schmitt22nat,Bange24SciAdv,Schmitt25natpho}, respectively. 
In short, the multidimensional photoemission data are acquired with a time of flight momentum microscope (ToF-MM, Surface Concept GmbH), and the extreme-ultraviolet light pulses (26.5$\,$eV, 20$\,$fs, $p$-polarized) are generated with a table-top high-harmonic-generation beamline driven by a 300$\,$W fiber laser (AFS Jena) operating at 500$\,$kHz~\cite{Keunecke20timeresolved, Merboldt25NatPhys}. 
The pump light (1.9$\,$eV, 40$\,$fs, $s$-polarized, 3$\,\upmu$J/cm$^{2}$ absorbed fluence, 9$\times10^{12}\,$cm$^{-2}$ exciton density) is generated by an optical parametric amplifier (Light Conversion). 
We measured energy- and in-plane momentum-resolved photoemission data cubes for selected pump-probe delays. 
The raw data collected by the instrument is the starting point for further processing and analysis with the presented software package.

\subsection{Energy calibration for ToF-MM}
When using a ToF-MM, the ToF of the electrons has to be converted to kinetic energy. 
This is achieved via an energy calibration. 
The mapping $E_{\rm eV} = f(\tau)$ that connects the measured ToF $\tau$ to an energy $E_{\rm eV}$ in eV, as for example given in~\cite{Spiecker1998_tof}, can be determined by recording a series of identical photoelectron spectra that are offset in ToF by a varying sample voltage $U_\text{sample}$. 
This results in spectra which are shifted in time of flight due to a known energy offset $E_\text{offset} = eU_\text{sample}$. 
By recording at least two of these spectra, one can approximate the mapping function $f(t_\tau)$ as a Taylor polynomial. 
The coefficients of this expansion can be determined by minimizing the variance between the recorded spectra when accounting for the offset introduced by the sample voltage. 
The \textit{energycal} tool of \momipy{} provides two methods of approximating the Taylor coefficients: first, by manually tuning the coefficients up to third order using the GUI tool until the spectra overlap. 
Second, given reasonable initial values, \textit{energycal} also provides a least-squares fit. 
The fit allows including higher-order coefficients, which may be necessary for photoelectron spectra spanning more than 10~eV.

\subsection{Momentum distortion and calibration}
Energy-dependent momentum distortions introduced by the electrostatic lenses can be characterized by using either a separate grid measurement, well-identifiable high-symmetry points in the Brillouin zone, or, if detected, the photoemission horizon. 
If the distortions can be approximated using an (invertible) affine transformation, the momentum coordinates are corrected with its inverse. 
In the same step, the data can be momentum-calibrated using the known momenta of the identified high-symmetry points or the radius from the photoemission horizon, which follows from the probe photon energy and the work function.

To quantify the distortions using a separate grid measurement, an isotropic, regular grid mask is inserted into the $k$/real space image plane, which is also projected onto the detector. 
Thus, deviations from the expected grid shape can be quantified. 
In \momipy{}, this is implemented by Fourier-transforming the ToF image slices of the grid measurement and identifying the position of the first-order grid peaks with respect to the zeroth order. 
The vectors from the zeroth- to first-order peaks give access to the scaling, shearing and rotation of the image. 
With that, one can quantify and correct the anisotropy of the image using an affine transform as well as scaling of the grid along the ToF axis. 
Such a grid measurement would also allow for nonlinear corrections of the rear lens system, but this is not yet implemented in \momipy{}. 
Note that, due to the position of the grid, this method only allows for correction of the rear projection lens system and not for the front lenses, which form the initial $k$-space and real-space images. 
However, most of the distortions originate from the projection lens system, so this limitation is not severe.

Instead of analyzing the distortions via a grid measurement, it is possible to quantify them using known landmarks in the momentum/real space image. 
Ref.~\cite{Schmitt25natpho} shows how this can be done in real space imaging mode. 
For momentum-space-specific analysis, \momipy{} offers isotropic and anisotropic momentum correction tools. 
They function by fitting an (an-)isotropic affine transform from source image points, e.g. high-symmetry points in the Brillouin zone whose position is determined from the measured data, to target points, e.g. theoretically calculated high-symmetry point positions. 
If this is performed for at least two ToF slices (kinetic energies), the affine transform parameters can be extrapolated to the full spectrum. An advantage of this correction approach is that it usually also calibrates the momentum axis, since the reciprocal space coordinates of the target features are known a priori.

It is also possible to calibrate the momentum axes and even to correct limited distortions based upon the photoemission horizon. Since the radius of the photoemission horizon $r_\mathrm{hor}$ follows 
\begin{equation*} 
r_\mathrm{hor}  = \sqrt{\frac{2m_e E_{\rm kin}}{\hbar^2}} = \sqrt{\frac{2m_e \left(h\nu - \Phi - E_\mathrm{B}\right)}{\hbar^2}},
\end{equation*} 
where $m_e$, $h\nu$, $\Phi$, and $E_\mathrm{B}$ are the electron mass, photon energy, work function and binding energy, respectively, this requires that the kinetic energy relative to the photoemission cutoff (or the photon energy, work function and binding energy) are known. 
Furthermore, clipping in the lens column must be avoided in order to measure the full photoemission horizon. 
If these conditions are satisfied, \momipy{} can be used to fitting the size of the horizon using edge detection and ellipse fitting. 
As this can be performed for each constant energy slice in the data, an energy-dependent momentum calibration can be extracted. Additionally, since the true photoemission horizon is circular, an anisotropic scaling can be corrected for.

\subsection{Space-charge correction}
To correctly interpret the binding energy of relevant features, one must ensure that the kinetic energy of the photoelectrons is not influenced by space-charge effects. 
Space-charge effects can be grouped into deterministic and non-deterministic phenomena~\cite{Schoenhense2015_ultramic}. 
If the deterministic effects can be quantified, they may also be corrected for. 
To this end, \momipy{} provides tools to identify and compensate for space-charge-induced momentum-dependent energy shifts using either specific equienergetic features or a space-charge-free reference spectrum. 
In principle, both approaches function in the same way: by extracting the momentum-dependent energy shift introduced by the space charge, which often follows a two-dimensional Lorentzian shape~\cite{schoenhense_space_charge}.
The space-charge shift then corrected by applying the inverse of this shift. 

The first correction method, using equienergetic features, determines the momentum-dependent energy shift from selected points in momentum space, such as the $\Sigma$ valleys in the bilayer \mos{} example. 
A 2D Lorentzian function is then fitted to the extracted energy shifts to approximate the space-charge profile. 
In general, the more samples are taken at different locations, the better the approximation. However, this requires knowing the (relative) energy of specific features, which limits the general applicability of this approach. 

Another approach is to measure a space-charge-free reference spectrum, for example, by blocking the pump beam or by reducing the probe flux.
From this, the momentum-dependent shift between the reference and space-charge-shifted spectra can be determined. 
This yields a 2D scalar field representing the space-charge-induced energy shift as a function of the momentum coordinates. 
A 2D Lorentzian is fitted to this scalar field, which again serves as the basis for quantification and correction. A possible correction can then be carried out by applying the inverse of the determined space-charge shift to the dataset.

\subsection{MMData}
The main data storage for multidimensional datasets in \momipy{} is the \texttt{MMData} class. 
It is built on top of the NeXus format (which in turn is based on HDF5) and thus inherently possesses some powerful capabilities. 
First, the data is automatically compressed. 
In the case of TIF stacks output from our end-station, this reduces the data size by a factor of roughly 10–100 without loss of information. 
Furthermore, HDF5 supports lazy-loading of datasets. 
This makes it possible to load only the parts of a file that are needed for evaluation, which is especially useful if datasets are very large or when working on machines with limited hardware. 
The loading API for \texttt{MMData} files also allows for binning of the data, which allows quick processing of large datasets at reduced energy, momentum, or time resolution.

\section*{Code availability}
\momipy{} is publicly available on \href{https://gitlab.gwdg.de/mathias-lab-public/momipy.git}{GitLab} under the \href{https://www.apache.org/licenses/LICENSE-2.0.html}{Apache License 2.0}.

\section*{Data availability}
No new data were generated in the context of this publication. 
The original, preprocessed data is published in Ref.~\cite{EKTPZZ_2026} and described in Ref.~\cite{werner2026_neq}. 
The raw instrument data along with the workflow is available on the GRO.data platform~\cite{momipy_gro_data}.

\section*{Acknowledgments}
This research was funded by the Deutsche Forschungsgemeinschaft (DFG, German Research Foundation) - Project Nos. 567214325, 566257456 and 535247173/SPP2244.
W.B. acknowledges support from the Dorothea Schlözer Postdoctoral Programme for Women. 
G.S.M.J. acknowledges support from the Netherlands Organisation for Scientific Research (NWO/OCW), as part of Quantum Limits (Project No. SUMMIT.1.1016).

\section*{Author contributions}
J.L., D.S., M.R., G.S.M.J. and S.M. conceived the research.
P.W., W.B., and G.S.M.J. wrote the code, documentation and examples.
P.W., W.B., and J.P.B. carried out the time-resolved momentum microscopy experiments.
P.W. analyzed the experimental data and prepared the figures.
P.W., W.B., G.S.M.J., and S.M. wrote the manuscript with contributions from all
authors. 

\section*{Disclosure statement}
No potential conflict of interest was reported by the authors.

\printbibliography
\end{document}